\documentstyle[aps,preprint,prl,12pt]{revtex}

\begin{document}

\title{\bf
Universal Behavior Of Lyapunov Exponents In Unstable
Systems  }

\author{Aldo Bonasera$^1$, Vito Latora$^{1,2}$ and
Andrea Rapisarda $^{2,3}$}

\address{\it  $^1$
Laboratorio Nazionale del Sud - I. N. F. N.,
via S. Sofia 44, I-95123 Catania, Italy    }

\address{ \it $^2$ Dipartimento di Fisica Universit\'a
 di Catania, c.so Italia 57, I-95129 Catania, Italy}

\address{\it $^3$ I. N. F. N. Sezione di Catania,
c.so Italia 57, I-95129 Catania, Italy}

\date{May, 4th 1995}
\maketitle


\begin{abstract}
We calculate the Lyapunov exponents in a classical molecular dynamics
framework.  The system is composed of few hundreds particles interacting
either through  Yukawa (Nuclear) or Slater-Kirkwood (Atomic) forces.   The
forces are chosen to give an  Equation of State that resembles the nuclear
and  the atomic $^4He$ Equation Of State  respectively near the critical
point for liquid-gas phase transition.  We find the largest fluctuations
for  an initial   "critical temperature".  The largest Lyapunov  exponents
$\lambda$ are always positive and can be very  well fitted near this
"critical temperature"  with a functional form  $\lambda\propto
|T-T_c|^{-\omega}$,  where the exponent $\omega=0.15$ is independent of the
system and mass number.    At smaller temperatures we find that
$\lambda\propto  T~ ^{0.4498}$, a universal behavior  characteristic of an
order to chaos transition.
\end{abstract}

\newpage
\noindent
The phenomena of phase transitions have been always a subject of great
interest  for many (generations of) physicists.  Very recently there has
been a large interest in  studying what happens when the system is not
composed of $10^{23}$ particles and confined  in a box but, on the
opposite, is composed of few hundred particles that are not confined.
This interest has born first in heavy ion collisions where one would like
to study the  Nuclear Equation of State (NEOS)\cite{ber84}.  In the nuclear
case it is not possible to study the  infinite number of particles case
(apart in stars with the obvious difficulties), so one  feasible way is to
perform proton-nucleus or nucleus-nucleus collisions.  In this manner  it
is possible to heat up and to explore different densities of the nuclei.
Of course the  problem is complicated apart from the fact of having a small
number of particles, also  from the presence of Coulomb force, angular
momentum and from the approximate  knowledge of the "thermodynamical
parameters", i.e. temperature, density and pressure
\cite{gro92,lat94,gil94}.  Similar  problematic can be found in the study
of metallic clusters and/or fullerenes\cite{bon95}.
One of the first questions that arises is: does it make sense to speak
about phase  transitions for a system made of 100-400 particles?  To answer
this question the authors  of refs.\cite{lat94,bon95} and \cite{bel95} have
solved the exact classical many body problem (Classical Molecular
Dynamics-CMD) numerically for particles interacting through two body
forces.   In  particular preparing 100 particles in the ground state and
giving to the particles an  excitation energy (or temperature T) the
following results are found:

1) for a given initial temperature $T_c$  the system undergoes
multifragmentation \cite{lat94,bel95,eos}.   The resulting mass
distributions display a power law $Y(A)\propto A^{ -\tau}$ with
$\tau=2.23$, which  is  exactly the value predicted in the Fisher droplet
model for a system near the liquid-gas  phase transition point
\cite{hua87}.  Other critical exponents have also been estimated within the
same model \cite{eos}:  $\gamma=0.86$, $\beta=0.33$\cite{note}.  Similar
estimates but for experimental data on  Au+C collisions at 1 GeV/nucleon
\cite{gil94} give also reasonable values of all the exponents;

2) at the same temperature, fluctuations in the mass distribution are
maximum.  This is  revealed through the study on an event by event analysis
of intermittency signal, factorial moments and Campi's plots
\cite{bel95,cam86,plo90};

3) the "critical temperature" follows the scaling law $T_c/|E_b| =
constant$\cite{bon95}.  Where $E_b$ is  the  binding energy of the system
(10.5 MeV for nuclei, 9.5 eV for C60 fullerenes and  50.5 $^0K$ for $^4He$
molecules);

4) at  very small temperatures or, equivalently, small excitation energies,
the events are  typical evaporation like events, i.e. with a final mass
distribution composed of a big  fragment and many small ones : monomers,
dimers etc.

For some initial conditions the system displays large fluctuations, thus we
expect  that other indicators of fluctuations -the Lyapunov exponents -
should be positive.  The  values of the Lyapunov exponents for systems
undergoing a phase transition are not  known (yet).  In particular the
relationship between thermodynamical and dynamical  properties have not
been exhaustively explored.  Ours is the first estimate of the
Lyapunov exponents for systems exhibiting a critical behavior (possibly
because of  a  liquid-gas phase transition).  In a previous exploratory
work, but for a two dimensional  system and in the mean-field
approximation, the largest  Lyapunov exponents were calculated in the
spinodal region\cite{ba95}.  It is well known, however, that the mean-field
approximation  gives a rough estimate of the critical
exponents\cite{hua87},  thus it is interesting to estimate their  values in
CMD.

We can summarize the main results of this work as follows:

\noindent
independently of
the studied system, i.e. nuclei, helium molecules, the largest  Lyapunov
exponents satisfy, similarly to the Landau theory of phase transitions, the
relation
\begin{equation}
  \lambda~ \propto ~ | T - T_c | ^{-\omega},   ~~~~~~~~T\sim T_c
\end{equation}

\noindent
where $\omega=0.15\pm0.04$.

At smaller temperature the Lyapunov exponents scale as

\begin{equation}
  \lambda ~ \propto ~  T  ^{ln2\over {ln\delta}},
\end{equation}

\noindent
with $\delta=4.669...$ the Feigenbaum constant \cite{fei80} which
indicates a typical transition from order to chaos\cite{hub80,chaos}.

Our studies are based on direct numerical simulation of an expanding system
in  classical molecular dynamics.  In particular we have studied a
"nuclear" system where  "neutrons" and "protons" interact through Yukawa
potentials.  Details of the forces can  be found in
refs.\cite{lat94,len90}.  The parameters entering the potentials have been
fitted in such a  way to have a ground state density of $0.16~ fm^{-3}$, a
binding energy of -16 MeV/nucleon  for an infinite system.  For finite
systems the binding energy is -10.5 MeV/nucleon  without Coulomb.  In this
work we will discuss the parameter set that gives a soft EOS,  i.e.
compressibility K=200 MeV \cite{len90} and the Coulomb interaction will be
neglected.  We notice that the use of the set  corresponding to a stiff
EOS, K=380 MeV, gives the same results.

We have also studied He atoms interacting through the potential\cite{sla31}
\begin{equation}
v(r)= 5.67 \cdot 10^6 ~ e^{-21.5{(r/\sigma)}} - 1.08 ~({\sigma\over r})^6
{}~,
\end{equation}

\noindent
where $\sigma=4.64 $ Angstrom and $v(r)$ is  in Kelvins.

In order to calculate the largest Lyapunov exponents \cite{chaos}
 we first define the following metric in phase-space
\begin{equation}
  d(t) =   ~ \sqrt{ \sum_{i=1}^N  [ \alpha ~ ({\bf r'}_1(t) -
{\bf r'}_2(t))^2  +
\beta ~ ({\bf p'}_1(t)) - {\bf p'}_2(t))^2 ]_i }
           ~  ~ ,
\end{equation}
where the sum runs over all the $N$ particles of the system, the subscript
1 and 2 refers to  two events that at time t=0 differ of an infinitesimal
quantity $d(0)=10^{-6}$ or less.  The ${\bf r'}$  and ${\bf p'}$ are scaled
positions and momenta.  In our case
\begin{equation}
        {\bf r'}(t)={\bf r}(t)/R_{rms}  ~ ,
\end{equation}

\begin{equation}
        {\bf p'}(t)={\bf p}(t)/P   ~ ,
\end{equation}

\noindent
where $R_{rms}$ is the root mean square radius and $P=\sqrt{2 m T}$ is an
average momentum, being  m the nucleon (or the atom) mass and T the initial
temperature (see below).  Normally the Lyapunov  exponents are calculated
for systems bound in phase-space.  This is not always the case  in our
simulations since for high excitation energy the phase-space explodes.   In
order to  be certain of the soundness of our results we calculated the
metric for three cases:  1) $\alpha$=1  and $\beta$=1; 2) $\alpha$=1 and
$\beta$=0;  3) $\alpha$=0 and $\beta$=1.   The results are independent on
the choice of  the metric, cases 1-3, as it should be.

The Lyapunov exponents $\lambda$ are obtained from the relation
\begin{equation}
        d(t)=d(0) ~ e^{\lambda t}   ~ .
\end{equation}
In our numerical simulations we prepared a system of 100 particles in their
ground state.   Then we distributed the momenta of the particles according
to a Maxwell-Boltzmann  distribution at temperature T and let the system
evolve in time by solving the classical equations of motion.  At each
temperature 100 events were generated.  For each  event (test event) ten
other events were generated, each event differing from the test event  of
d(0).  The exponents were obtained by averaging over all the events.  In
order to get $\lambda$  numerically, we  calculated the ratio $d(t)/d(0)$
and fitted its exponential growth. As a check we estimated also
\begin{equation}
        \dot d(t)/d(t)=\lambda ~ .
\end{equation}
In figure 1 we plot typical evolutions of $d(t)/d(0)$
at three temperatures T=2, 5  and 20 MeV
for the nuclear system.  We see that the distance increases exponentially
with time and can be very well fitted with a straight line (in a semilog
plot) whose slope is just the Lyapunov exponent.
The same results are found by using eq.(8).   In general we get Lyapunov
exponents which are always positive  at all temperatures and have a maximum
value   at  $T\sim 4.5~  MeV$.   A similar behavior is found for the
atomic case with a maximum at $T\sim 21.8~ ^0 K$.
In order to compare such different systems like nuclei or
atomic He, we scale  temperatures and times with typical values of the
different systems.  Usually one scales  with the values obtained at the
critical point.  Since we do not know these values a priori,  we scale the
temperature with the absolute value of the binding energy and the (inverse)
times with a typical value $\lambda_0$
\begin{equation}
        \lambda_0 = \sqrt{2 |E_b| /m} /R_0  ~ .
\end{equation}

Recall that in order to derive the EOS \cite{hua87} one needs to know the
hard core radius and the  depth of the two body potential.  These
quantities are proportional to the average  equilibrium distance between
particles           $R_0 = R/A^{1\over3}$, - where R is the radius of the
system  and A the mass number -  and to the binding energy.  In our case
$\lambda_0 = 0.155~ c/fm $ for the nuclear case  and $\lambda_0 = 8.82
\cdot 10^{-5} ~\alpha c /a_0$ for the atomic one, being $\alpha$ the
fine structure constant, $c $  the speed of
light and $a_0$ the Bohr radius. In figure 2 we plot the Lyapunov exponents
at each initial temperature (in units  of their typical values) for the
nuclear (circle symbols) and atomic cases (squares).  Note  that the
maximum value of $\lambda$ is obtained at the initial scaled temperature
\begin{equation}
        T_c/|E_b|\sim0.43
\end{equation}
\noindent
for both systems.  For such temperatures the systems undergo
multifragmentation and a power law  in the mass distribution and  factorial
moments is found \cite{lat94,bel95}.  Thus our result confirms that in this
temperature region  fluctuations are largest.   Inspired by the Landau
theory of phase transition, we  parametrized the exponents according to
eq.(1). The fit is also
displayed in figure 2 (full and dashed curves).  The same value
$\omega=0.15\pm0.04$ was used in both cases.
Note the good agreement with the scaled
$\lambda$ at all  temperatures but the lowest ones.  For such small
temperatures there is not any  multifragmentation of the system, indeed the
mass distribution reveals a typical cases of  evaporation.  Using the same
language as in the theory of phase transitions we could say  that the small
temperature cases explore densities and pressures outside the spinodal
region.  Note that the  absolute values of the scaled exponents differ of
less then $30 \%$ for the two cases.
Correspondingly the two fits differs only for a multiplicative constant $C$.
The  reason for such a small
discrepancy is due to the fact that in the nuclear case we have two
fluids, neutrons and protons.  We will discuss this point in more detail in
a following  publication\cite{blr}.  We also stress that these results are
independent on the chosen metric,  i.e. in the values of $\alpha$ and
$\beta$ in eq.(4).  We have also tested the results by changing the  mass
of the fragmenting system.  The Lyapunov exponents remained the same for
masses ranging from 50 to 400 particles\cite{blr}. The uncertainties
reported in figure 2 are of the  order of $\sim 10\%$.

A scaling law of Lyapunov exponents of the kind
\begin{equation}
 \lambda \propto (A - A^*)^{ln2/ln\delta} = (A - A^*)^{0.4498}  ~ ,
\end{equation}
where A is a control parameter  and $\delta=4.669...$ the Feigenbaum
constant \cite{fei80}, is typical of order to chaos transitions. The value
$A^*$ is the critical value which indicates the onset of chaos. The
expression (11) was initially found in the logistic  map \cite{hub80}, but
various experiments have confirmed its general validity \cite{chaos}. In
our case the largest Lyapunov exponents $\lambda$  are positive for all
finite temperatures,i.e. the dynamics  is always chaotic, but they tend to
zero as the temperature T goes to zero.  Actually for T=0 the systems are
frozen in their ground state and $\lambda$ vanishes.
Thus chaos starts at $T>0$.

In figure 3  we plot a magnification of figure 2 for very small
temperatures. The full curve is the expression (11) multiplied by a
constant fitted on  the  numerical points.  In this case the control
parameter is the scaled temperature  and $A^*$  corresponds to T=0. The
agreement is really impressing for both systems up to $T/|E_b|\sim 0.1$.

We can try to understand this behavior by recasting our microscopic
dynamics in terms of a phenomenological macroscopic model.  At very low
temperatures one gets essentially evaporation events. Considering a
discrete map, at time $n$ the system evaporates $z_n$ particles and it will
continue evaporating  $q z_n$ particles until the excitation energy is
zero.   The number of  evaporated particles at time n+1 is thus
$z_{n+1}=(1+q) z_n$.   If $ z_0$ is the final number of  particles  and  we
assume that  the number of evaporated particles decreases linearly with
time because the excitation energy is decreasing, we obtain $z_{n+1}= (1+q)
z_n (1-z_n/z_0)$.  This is nothing else that the logistic map if  $1+q=A$
and $x_n=z_n/z_0$.

Thus we have two different mechanisms at play.   The first one for small
temperatures gives a transition from order (the ground state) to chaos and
has a dynamical  origin.  The second mechanism, for  reduced temperatures
larger then 0.1 has a  thermodynamical origin appropriate for a second
order phase transition.  Loosely speaking  we have given evidence for
"critical chaos" in the latter case.

In conclusion, in this work we have calculated the  largest Lyapunov
exponents as a function of the initial temperature for an  expanding
system composed by 100 particles in the framework of classical molecular
dynamics.   We have shown that these exponents are always positive and
have their largest value  at a temperature of $\sim 4.5~ MeV$ for the
nuclear case and $\sim 21.8 ~^0K$ for the atomic one.   We  have also
demonstrated that the $\lambda$ obey to universal scaling  laws. They
fulfill the relation  $|T-T_c|^{-\omega}$,  $\omega=0.15 \pm 0.04$,
similarly to the  Landau theory of phase transitions near the critical
point.  At the same time for smaller temperatures (evaporation events)
Lyapunov exponents exhibit a general transition from  order ($T=0$) to
chaos ($T>0$). We feel that further  investigations following the ideas
presented in this paper may help our understanding of  order and disorder
in classical systems and, after all, in (part of) Nature  itself.

\bigskip
\bigskip
\centerline{\bf  Acknowledgements}

We thank Takayuki Kubo for providing the CMD code for He atoms.
Stimulating  discussions with M. Baldo, M. Belkacem, G.F. Burgio and
V. Kondratiev are  acknowledged.

\begin{figure}
\noindent
\caption{ The ratio $d(t)/d(0)$ is plotted  as a function of time at three
initial temperatures $T=2,5,20~MeV$ for the nuclear system. The
dashed lines are fits whose slope give the typical $\lambda$
for these temperatures after averaging over hundreds of events. }
\end{figure}

\begin{figure}
\noindent
\caption{The scaled largest Lyapunov exponents $\lambda/\lambda_0$
are plotted vs. the scaled initial temperature $T/|E_b|$ for the nuclear
(circles) and  the atomic (squares) case. The full and dashed lines are fits
obtained with the functional form $C |T-T_c|^{-\omega}$ where
$\omega=0.15\pm 0.04$. The parameters of the fits are C=0.25 and C=0.3
for the nuclear and atomic case respectively.
See text for further details.}
\end{figure}

\begin{figure}
\noindent
\caption{A magnification of figure 2 at very small temperatures.
The full curve is a fit with the functional form   $K ~ T~^{0.4498}$
where $K=0.55$ is the fitted parameter.
See text for further details.}
\end{figure}


\begin{thebibliography}{99}
\bibitem{note}
Note that the value of $\gamma$ is smaller than the accepted value
$(1.3\pm 0.1)$\cite{hua87}.  We stress the  difficulty to extract such an
exponent from the simulation since one needs events on the  liquid side
where multiplicity is small.   In order to see that, we estimate other
exponents  using the well known relations: $\tau=2 + 1/\delta$, which gives
$\delta=4.35$,  but from $\gamma=\beta(\delta-1)$ we obtain $\gamma=1.1$
which is closer to the experimental value and larger then  the direct
estimate.
\end{thebibliography}
\end{document}